\documentclass[fleqn,usenatbib]{mnras}

\usepackage[T1]{fontenc}

\DeclareRobustCommand{\VAN}[3]{#2}
\let\VANthebibliography\thebibliography
\def\thebibliography{\DeclareRobustCommand{\VAN}[3]{##3}\VANthebibliography}

\usepackage{graphicx}	
\usepackage{amsmath}	
\usepackage{amssymb}	
\usepackage{threeparttable}

\newcommand{\hii}{H\,{\sc ii}}
\newcommand{\hi}{H\,{\sc i}}

\newcommand{\siv}{[S\,{\sc iv}]}
\newcommand{\bralpha}{Br\,$\alpha$} 
\newcommand{\brgamma}{Br\,$\gamma$} 
\newcommand{\mbralpha}{{\rm Br}\,\alpha}
\newcommand{\halpha}{H\,$\alpha$}

\newcommand{\hei}{He\,{\sc i}}
\newcommand{\msun}{$\rm M_\odot$}
\newcommand{\lsun}{$\rm L_\odot$}

\newcommand{\kms}{$\rm km\,s^{-1}$}
\newcommand{\varcs}{\kms~$\rm arcsec^{-1}$}
\newcommand{\vpc}{\kms~$\rm pc^{-1}$}

\title[Spectroscopy of NGC 1569's embedded {\hii} region]{Observing the influence of the youngest super star clusters in NGC~1569: Keck Brackett $\alpha$ spectroscopy}

\author[D. P. Cohen et al.]{
Daniel P. Cohen,$^{1}$\thanks{E-mail: \href{mailto:daniel.parke.cohen@gmail.com}{daniel.parke.cohen@gmail.com} (DPC)}
Jean L. Turner,$^{1}$
Sara C. Beck,$^{2}$ 
S. Michelle Consiglio,$^{1}$
\\
$^{1}$Department of Physics and Astronomy, University of California, Los Angeles, CA 90095-1547, USA\\
$^{2}$School of Physics and Astronomy, Tel Aviv University, Ramat Aviv, Israel
}

\date{Accepted 2021 Feb 26. Received 2021 Feb 26; in original form 2021 Nov 16}

\pubyear{2021}

\begin{document}
\label{firstpage}
\pagerange{\pageref{firstpage}--\pageref{lastpage}}
\maketitle

\begin{abstract}

We report Keck-NIRSPEC observations of the Brackett $\alpha$ 4.05 $\mu$m recombination line across the two candidate embedded super star clusters (SSCs) in NGC~1569. These SSCs power a bright {\hii} region and have been previously detected as radio and mid-infrared sources. Supplemented with high resolution VLA mapping of the radio continuum along with IRTF-TEXES spectroscopy of the {\siv} 10.5$\mu$m line, the Brackett $\alpha$ spectra data provide new insight into the dynamical state of gas ionized by these forming massive clusters. NIR sources detected in 2$\mu$m images from the Slit-viewing Camera are matched with GAIA sources to obtain accurate celestial coordinates and slit positions to within $\sim0\farcs1$. {\bralpha} is detected as a strong emission peak powered by the less luminous infrared source, MIR1 ($L_{\rm IR}\sim 2\times10^7$~{\lsun}). The second candidate SSC MIR2 is more luminous ($L_{\mathrm{IR}}\gtrsim4\times10^8$~\lsun) but exhibits weak radio continuum and \bralpha\ emission, suggesting the ionized gas is extremely dense ($n_e\gtrsim 10^5$ cm$^{-3}$), corresponding to hypercompact {\hii} regions around newborn massive stars. The {\bralpha} and {\siv} lines across the region are both remarkably symmetric and extremely narrow, with observed line widths $\Delta v \simeq 40$ {\kms}, FWHM. This result is the first clear evidence that feedback from NGC~1569's youngest giant clusters is currently incapable of rapid gas dispersal, consistent with the emerging theoretical paradigm in the formation of giant star clusters.
 
\end{abstract}




\section{Introduction} \label{sec:intro}

As the birth places for nearly all massive stars, giant star clusters fundamentally shape the dynamics and chemistry of their host galaxies. Galactic observations of ancient globular clusters (GCs) give a direct view into the evolved phase of massive star cluster evolution. These investigations have revealed unexpected properties of GCs that are determined entirely within their {\it earliest} evolutionary phase. Of particular interest are how massive clusters survive the first $\sim5$ Myrs of intense feedback from massive stellar evolution and remain bound for Gyrs, and how they form the multiple stellar populations that are ubiquitous in GCs \citep[e.g.,][]{piotto2015,bastian2018}. The only young clusters with comparable masses  exist within rare {\it extragalactic} starburst regions. Super star clusters (SSCs) host $10^5$-$10^6$~{\msun} in stars and gas and  may represent  directly observable analogs of proto-globular clusters \citep[e.g.,][]{turner2009}. 

The youngest SSCs are still deeply embedded within their natal clouds of gas and dust and suffer large internal extinction, rendering them unobservable in optical-NIR wavelengths. However, these clusters ionize luminous compact {\hii} regions that are detected as radio-IR ``supernebulae". The dynamics of such supernebulae reflect the collective influence of rapidly evolving massive stars \citep{beck2008}, providing unparalleled insight into the mechanisms of retention or expulsion, hence survivability, of massive clusters like GCs \citep{elmegreen2018,gray2019}. Requiring high spatial and spectral resolution in the IR/radio, direct measurements of supernebula kinematics have, until recently, been largely infeasible.  

Observations of SSCs made within the last few years provide evidence for a fundamental difference in how feedback operates in the densest, most-massive star clusters. The supernebula in dwarf starburst galaxy NGC~5253 exhibits a recombination linewidth of merely $\Delta v\sim75$ {\kms}, similar to linewidths of Galactic HII regions, despite being powered by a $M\sim4\times10^5$~{\msun} cluster with $\sim$2000 O stars and hot molecular cores within its $r\sim$3 pc extent \citep{turner2000,meier2002,turner2003,turner2004,calzetti2015,smith2016,consiglio2017,turner2017,cohen2018,silich2020}.
Simulations find that the most highly concentrated forming SSCs host ionized stellar wind regions that can stall in their expansion due to critical radiative cooling, preventing a coherent cluster wind and allowing the cluster to retain enriched fuel for ongoing star formation \citep{silich2018,silich2017,gray2019,silich2020}. 
Finding and studying more SSCs such as the NGC~5253 supernebula is critical to solving the unanswered questions of GC formation and galaxy evolution. 

Perhaps the next best candidate of a proto-globular cluster after NGC~5253 is found within dwarf starburst NGC~1569, located at a distance of $D=3.4$ Mpc \citep{grocholski2012} and known for its two bright, visible SSCs, A and B \citep[e.g.,][]{hunter2000}. Currently NGC~1569's most active star formation is concentrated to an obscured region characterized by high extinction  \citep{galliano2003,pasquali2011,lianou2014,westcott2018}. Intense thermal radio continuum coinciding with bright {\hei} and Pa~$\beta$ emission indicates the presence of {\hii} regions excited by forming massive star clusters \citep{greve2002,clark2013}. The brightest {\hii} region, identified as {\hii} 2 in \citet{waller1991}, is located near a group of CO clouds at the end of a H$\alpha$ filament extending $\sim$2 kpc \citep{taylor1999}. {\hi} along this filament exhibits redshifted non-circular motion \citep{johnson2012}, indicating a potential inflow into the embedded star-forming region. This could be an example of filament-fed star formation, much like the case suggested for NGC~5253 \citep{turner2015,consiglio2017}.

 Two candidate embedded forming massive clusters in NGC~1569, identified with high resolution {\siv} 10.5 $\mu$m and mid-infrared continuum imaging in \citet{tokura2006} as MIR1 and MIR2, may be the very youngest clusters in NGC 1569. They appear to be responsible for ionizing {\hii} 2.  \citet{tokura2006} find that the brighter source MIR1 has an estimated intrinsic infrared luminosity on the order of $10^7~\rm L_\odot$, corresponding to $\gtrsim$100 O7 stars (revised for $D=3.4$ Mpc). Despite a weaker line flux, the second source MIR2 is estimated to have a higher intrinsic luminosity of $\gtrsim 10^8~\rm L_\odot$. MIR2 should be easily detectable in the radio continuum, but exhibits no clear radio counterpart. The IR luminosity of MIR2 is consistent with ionization from thousands O7 stars. This suggests MIR2 is a SSC in the earliest phases of formation, where the density is so high that the nebula is optically thick in free-free emission and has a bright and hot dust  continuum emission.  MIR1 and MIR2 are separated by only 1.5\arcsec, or 30 pc in projection. Their relation along with their gas kinematics, is unclear.

In this paper we present the {\bralpha} emission line spectrum across embedded SSC candidates MIR1 and MIR2 and the brightest {\hii} region in NGC~1569, obtained with the NIRSPEC echelle spectrograph on Keck. At 4 $\mu$m, the {\bralpha} line is less effected by extinction than {\halpha} or {\brgamma}. The NIRSPEC observations are complemented with high resolution VLA mapping of the 6 cm continuum along with high resolution spectroscopy of the {\siv} 10.5 $\mu$m line from the IRTF. For the first time, the spectra reveal the dynamical nature of the embedded ionized gas in the giant {\hii} region at a $\sim$10 {\kms} resolution.
The adopted distance of of $D=3.4$ Mpc for NGC~1569 \citep{grocholski2012} corresponds to a scale of 1$\arcsec$=16 pc.

\section{Observations \label{sec:observations}}

\subsection{Radio continuum with the VLA}\label{sec:observations:1}

 Radio continuum maps were used to position the NIRSPEC slit on NGC~1569's embedded giant {\hii} region complex and provide a powerful complement to the {\bralpha} spectroscopy. NGC~1569 was observed with the VLA in A configuration on 23 August 1999 at C (6 cm) band in program AT227 
(P.I. J. Turner). Fluxes were calibrated using 3C286; the absolute flux calibration is estimated to be to better than 5\%, although this does not include flux lost due to undersampling in these high resolution maps. The uvdata were tapered to give a beam of size $0\farcs78\times0\farcs66$ (FWHM) at P.A. = 71\degr. The shortest baseline is $\sim 10^4~\rm k\lambda,$ so structures of extent greater than $\sim$15\arcsec\ are poorly
sampled. The rms in the map is 0.12 mJy/beam. 

\begin{figure}
\centering
\includegraphics[width=0.99\columnwidth]{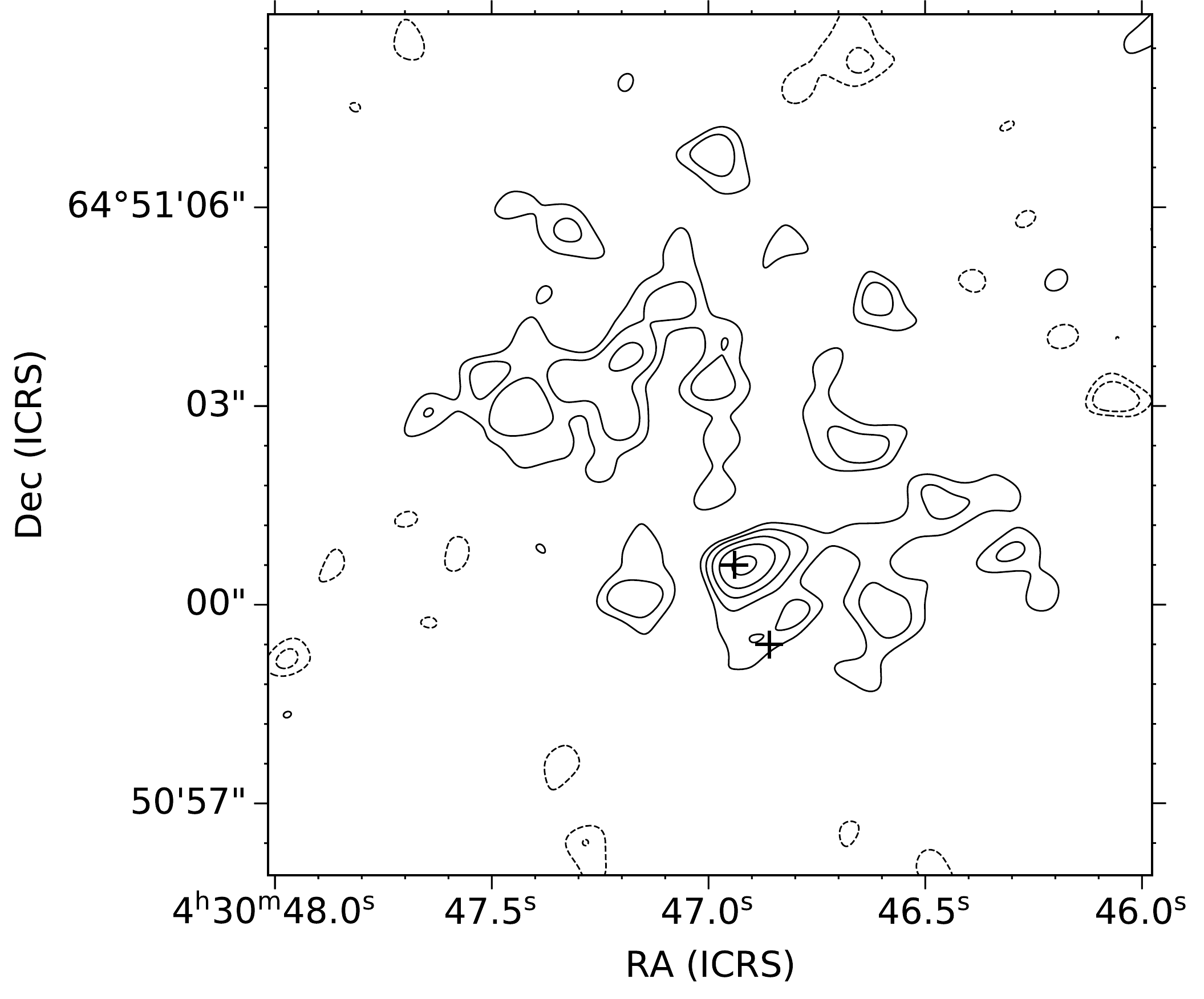}
\caption{VLA map of the 6cm radio continuum near the embedded {\hii} complex in NGC~1569. Contours are drawn at brightness levels $0.25~{\rm mJy/beam} \times 2^{n/2}$ for $n=\{0,1,2,3,4,5,6,7,8\}$, and dashed contours mark negative levels $n=\{0,1,3,4\}$. The peak brightness is 1.0911 mJy/beam. Crosshairs mark the positions of MIR1 and MIR2 from \citet{tokura2006}, who assumed MIR1 is coincident with the radio peak; these sources are also marked in Fig.~\ref{fig2}.
\label{fig1}}
\end{figure}

The 6~cm continuum map is shown in Fig.~\ref{fig1}. 
In lower resolution maps by \citet{lisenfeld2004} and \citet{westcott2018}, radio continuum emission is observed over a 2{\arcmin} region centered on
the visible SSCs in NGC 1569. The region mapped here is near their "{\hii}~2" source, which is identified
as radio source "M-1" in 1.4 GHz MERLIN maps of \citet{greve2002}. The radio continuum in this region is characterized by a flat spectral index \citep{greve2002,lisenfeld2004,westcott2018}, consistent with thermal free-free
emission. Extended emission in this smaller M-1 region is not well sampled by the high resolution VLA image. The total 6~cm flux for the region shown in the figure is roughly 20 mJy.

The compact radio source in Fig.~\ref{fig1} is source MIR1 identified by \citet{tokura2006}.
A Gaussian fit to the 6~cm 
source gives a position for this radio source of R.A., Dec. 
(ICRS) = 04$^{\rm h}$30$^{\rm m}$46\fs913$\pm$0\fs007,
 64\degr51\arcmin00\farcs57$\pm$0\farcs03, in good agreement with the 1.4 GHz MERLIN
 coordinates of \citet{greve2002}. The source appears to be slightly extended, with a peak
 flux of 1.3$\pm$0.1 mJy/beam and a total flux density of 1.9 mJy.
 The region corresponding to MIR2 has weak if any 6~cm continuum emission, and we put an upper limit of 0.3~mJy for a radio source at this location.

\subsection{{\bralpha} spectroscopy with NIRSPEC}\label{sec:observations:2}

\begin{figure*}
\centering
\includegraphics[width=0.95\textwidth]{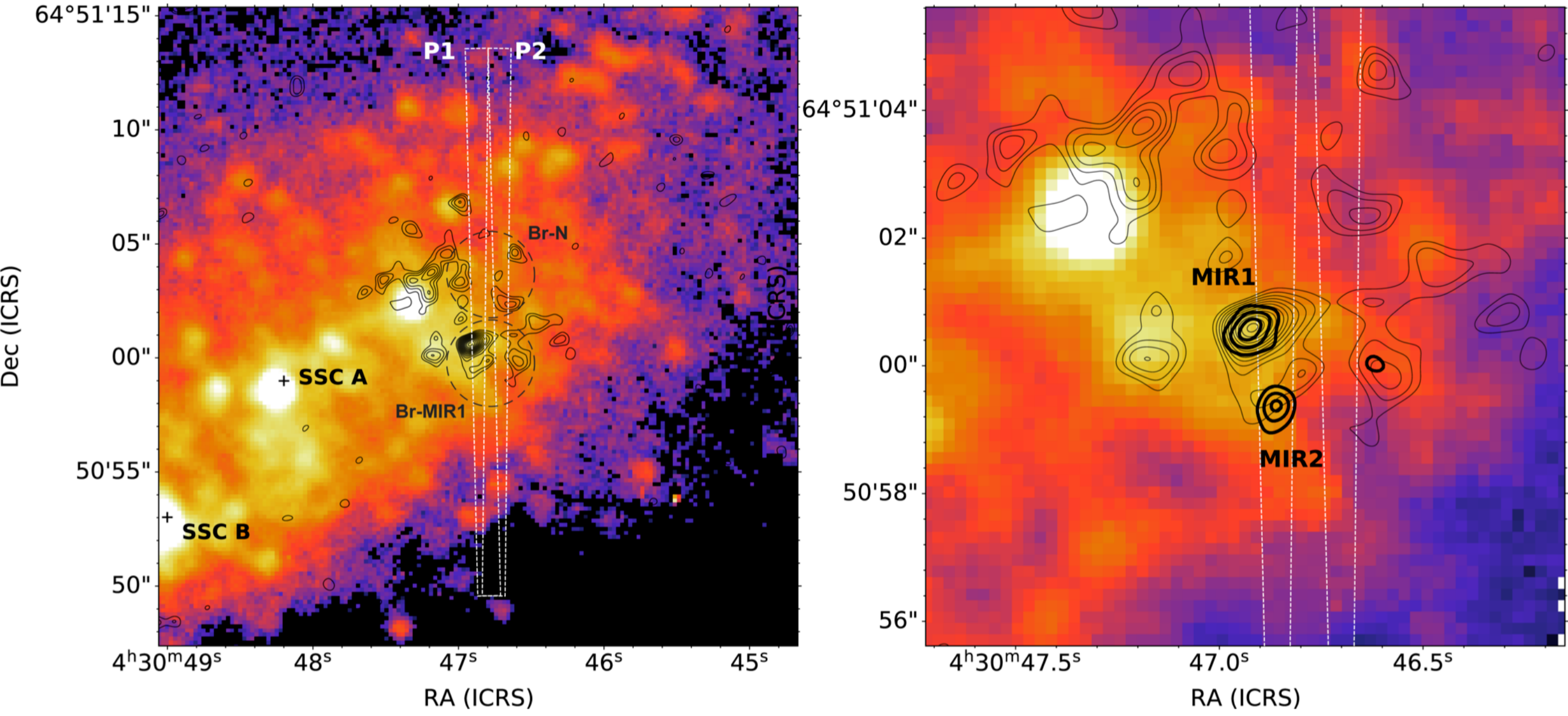}
\caption{ NIRSPEC observations of the embedded star-forming region in NGC~1569. {\it Left}) Positions of the $0\farcs432\times24{\arcsec}$ slits (white boxes) over the combined SCAM image of the KL-band continuum (colorscale). Slits were centered near the bright thermal radio continuum/MIR peak (Fig.~\ref{fig1}). The approximate locations of the two brightest {\bralpha} emission peaks Br-N and Br-MIR1 (Sec.~\ref{sec:results:1} and Fig.~\ref{fig3}) are marked by the black dashed circles. Optical SSCs A and B are shown for reference. \textit{Right}) Zoom-in of target region, showing the relation between 10$\mu$m continuum \citep[thick contours;][]{tokura2006}, radio continuum (thin contours), and KL-band (colorscale). MIR1 and MIR2 stand out as the brightest MIR sources in the region by far. The seeing measured in the SCAM imaging is roughly $1{\arcsec}\simeq16$ pc. \label{fig2}} 
\end{figure*}

 We observed the embedded star-forming region in NGC~1569 with NIRSPEC 
on Keck II \citep{mclean1998} during the first half-night on Dec 7, 2017.
Two high-resolution (echelle) spectra, each with 240s integration, were acquired in the KL band through a 0\farcs432$\times$24\arcsec\ slit positioned on the galaxy's thermal radio continuum peak and candidate embedded SSCs MIR1 and MIR2, as shown in Figure~\ref{fig2}. The echelle and cross-disperser angles were set to 64.42{\degr} and 34.3{\degr}, respectively, to observe the {\bralpha} line in the 19th echelle order. The resulting wavelength coverage of this order was 4.032-4.085 $\mu$m.  The calibration star, HD12365, was observed with an ABBA nod sequence prior to observations of NGC~1569. 

Each slit position yielded a 2D echelle spectrum which we rectified and reduced. Reduction involved the use of flat-field and ThArXeNe arc-lamp spectra acquired at the beginning of the night,
and a raw sky spectrum taken after NGC~1569 exposures at an offset of $\sim$50{\arcsec}. 
After first being cleaned of cosmic rays and bad pixels, raw science spectra were sky-subtracted (or $A-B$ subtracted in the case of HD12365) and divided by a median-normalized flat-field image. The spectra were then rectified -- mapped from the curved echelle order onto a grid with orthogonal spatial and dispersion axes, using transformations generated with the \texttt{SPATMAP} and \texttt{SPECMAP} routines from the \texttt{REDSPEC} reduction code\footnote{\url{https://www2.keck.hawaii.edu/inst/nirspec/redspec.html}}. The wavelength calibration was determined as part of the \texttt{SPECMAP} procedure by identifying arc-lamp lines and fitting their positions with a 3rd-order polynomial. 

\begin{figure*}
\includegraphics[width=0.75\textwidth]{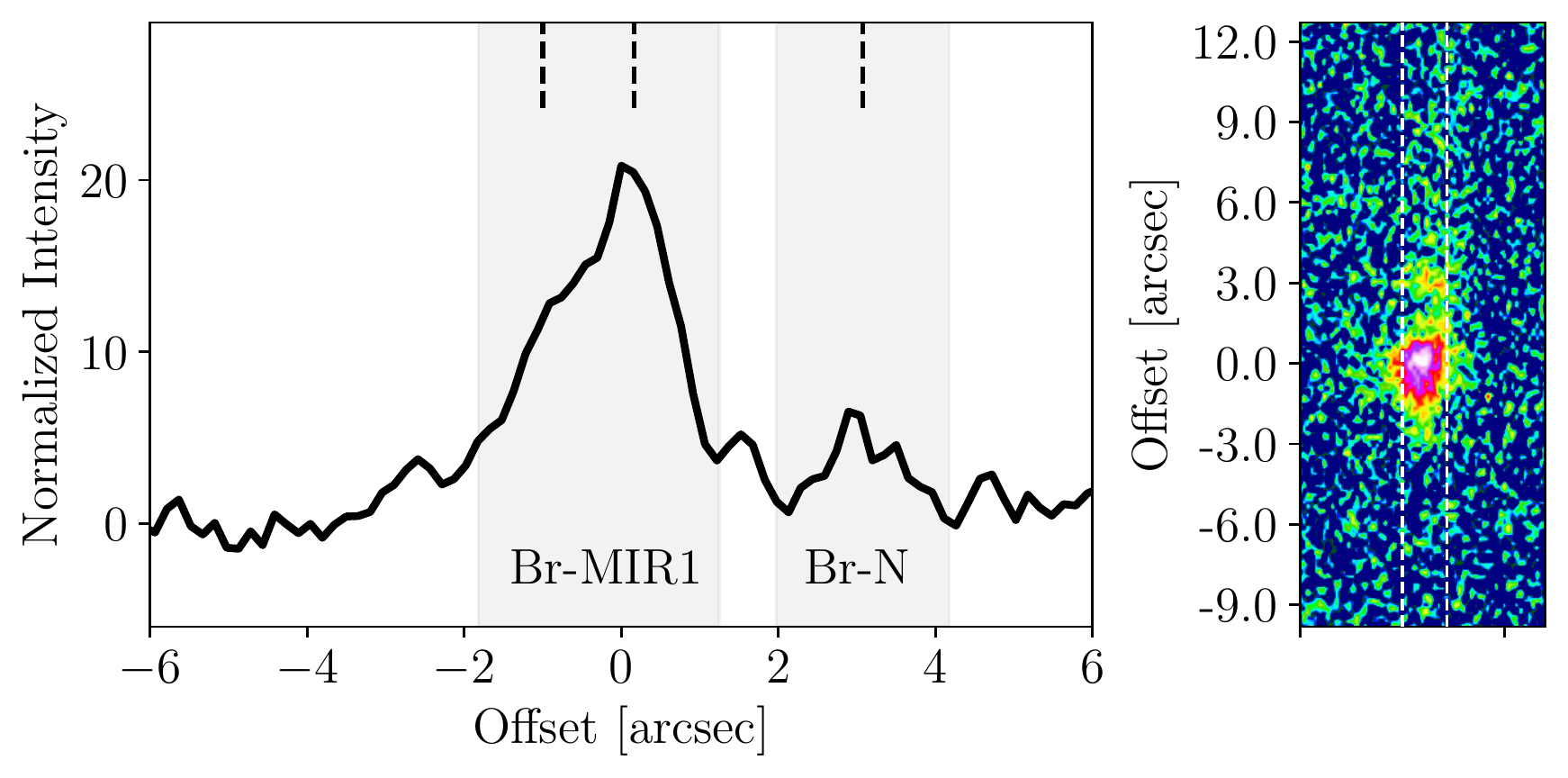}
\centering
\caption{The combined 2D echelle spectrum of {\bralpha} from NIRSPEC (right) and the extracted intensity profile along the slit (left). In the echelle spectrum, the spectral and spatial axes are along $x$ and $y$, respectively. The 1D spatial profile was extracted from the spectrum by averaging columns between the white dashed lines. Grey bands identify the strongest {\bralpha} emission sources: (1) Br-MIR1 encompassing the target {\hii} region complex and MIR sources, and (2) an unrelated source Br-N located 3{\arcsec} to its north. Dashed vertical lines mark best-fit peak positions (Br-MIR1 is fit with two blended components).  \label{fig3}}
\end{figure*}

\begin{figure}
\includegraphics[width=0.99\columnwidth]{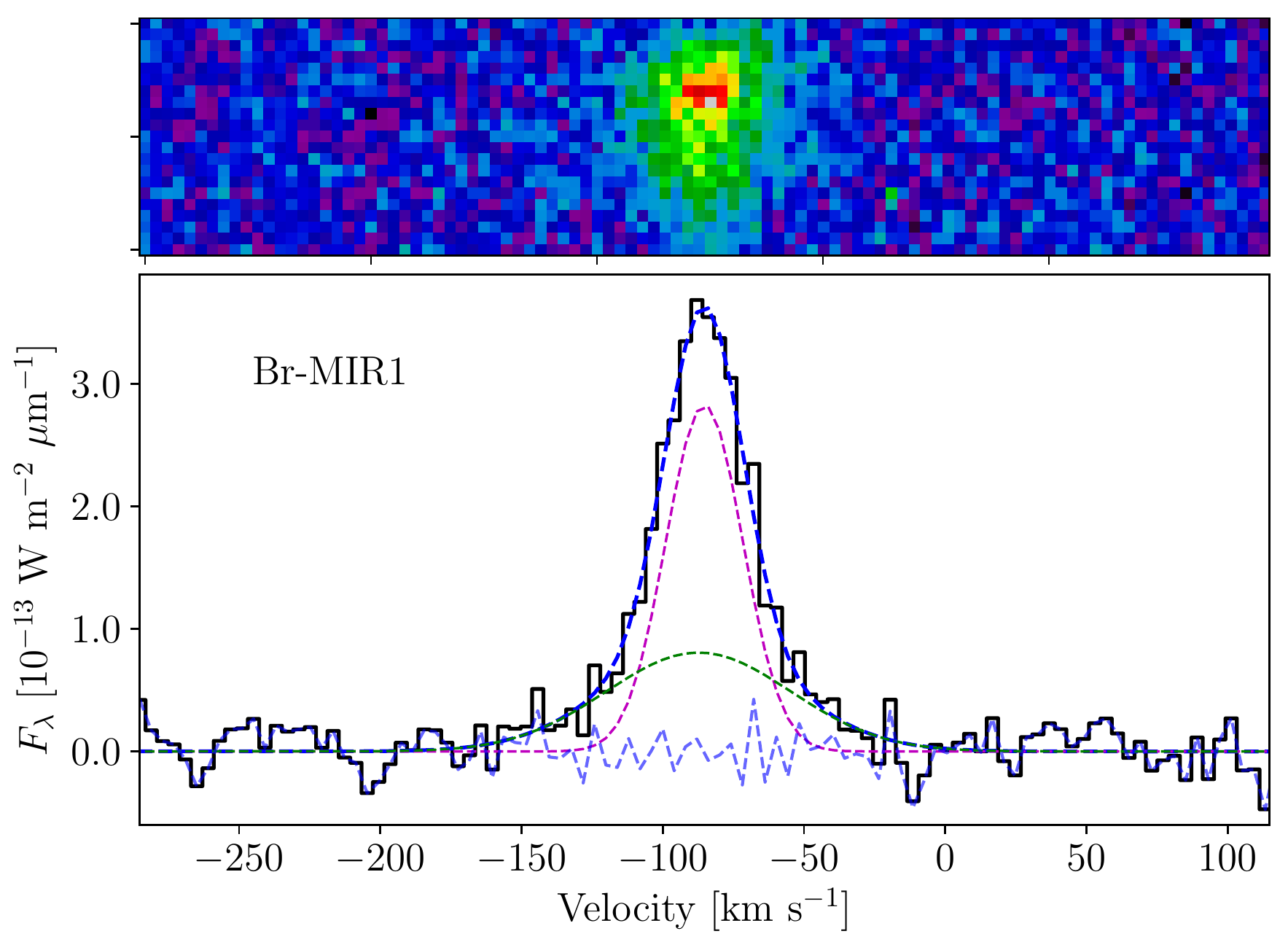}\vspace{3mm}
\includegraphics[width=0.99\columnwidth]{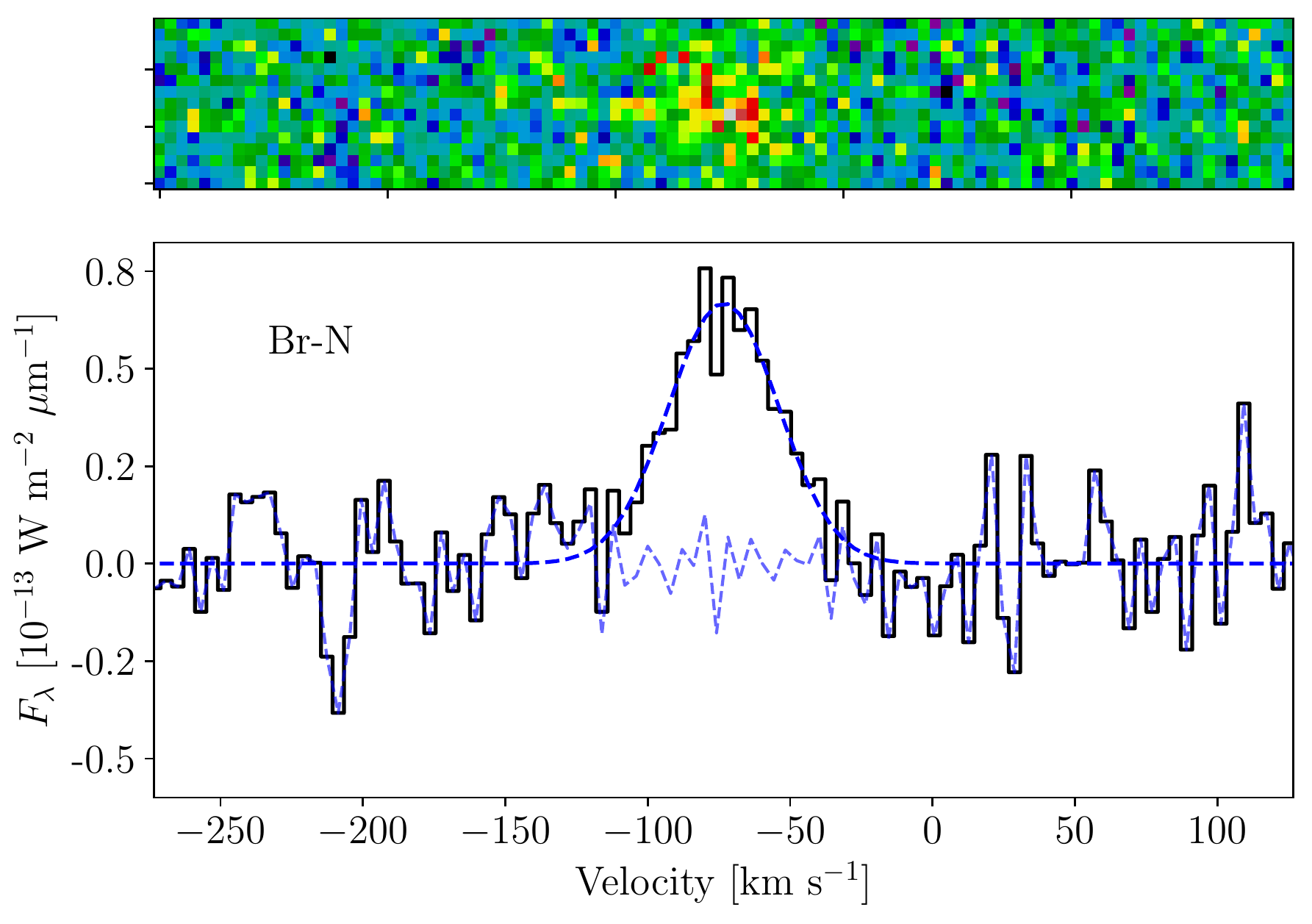}
\centering
\caption{ {\bralpha} line profiles of sources Br-MIR1 (top) and Br-N (bottom) identified in Fig.~\ref{fig3}, extracted by summing the rows in the 2D spectra (colorscale). Br-MIR1 and Br-N are detected with peak S/N ratios of $\sim$30 and 8, respectively. Dark and light blue dashed curves show the best-fit profiles and residuals. Two Gaussian components are fit to Br-MIR1 (magenta and green dashed curves), while Br-N is fit with a single Gaussian profile. \label{fig4}}
\end{figure}

Figure~\ref{fig3} shows the combined reduced echelle spectrum, along with the spatial profile of the Brackett $\alpha$ emission. The echellegram has spatial information along the vertical axis and spectral information along the horizontal axis, with a pixels size of $0\farcs152$ by $5.44\times10^{-5}$ $\mu$m. The instrumental spectral resolution at 4.051 $\mu$m is 12 {\kms}. The {\bralpha} line profiles of identified sources, reported in Figure~\ref{fig4}, are extracted from the echelle spectrum by collapsing along the spatial axis within apertures of size determined by source size. The photometric calibration was estimated with the spectrum of HD12365 by summing the star's spectral trace and adopting a 65\% flux loss by the $0\farcs5$ slit, allowing us to convert measured flux in counts s$^{-1}$ to physical units. Wavelengths and velocities quoted in this paper have been converted to the heliocentric frame.

\subsection{{\siv} spectroscopy with TEXES}\label{sec:observations:3}

\begin{figure}
\centering
\includegraphics[width=0.9\columnwidth]{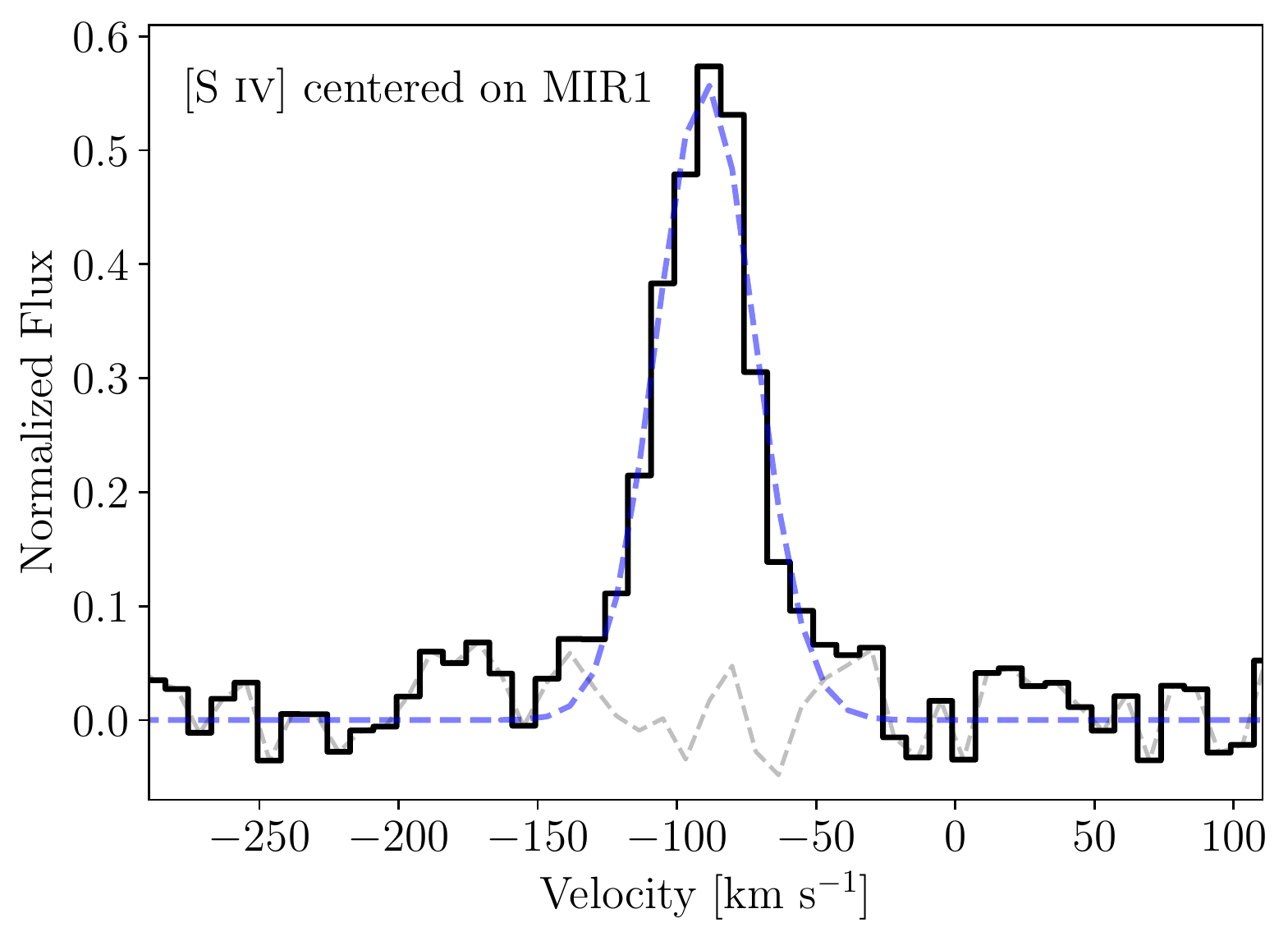}
\caption{{\siv} 10.5~$\mu$m line observed with TEXES at the position of the compact radio peak/MIR1 (black curve), along with the best-fit Gaussian profile and residuals (blue and grey dashed curves, respectively). 
\label{fig5}}
\end{figure}

 The {\siv} 10.5 $\mu$m line was observed across NGC~1569's radio cotinuum/MIR peak with TEXES \citep{lacy2002} on the NASA IRTF on Jan 1, 2012. The [SIV] line is excited by hot stars of more than $\sim 40~M_\odot$. It is affected by extinction approximately as much as is the Br~$\alpha$ and is less susceptible to thermal broadening. The 29{\arcsec} slit was oriented north-south and centered on the position of MIR1 using blind-offset pointing from a guide star. Additional spectra were acquired by stepping the slit 0\farcs7 east/west (half the seeing FWHM 1\farcs4). The instrumental resolution for the spectra (medium resolution mode) is $\sim$18 {\kms}. The {\siv} line profile observed at the central pointing is shown in Fig.~\ref{fig5} and discussed further in Sec.~\ref{sec:results:2}.

\section{Results} \label{sec:results}

\subsection{Relating  radio continuum and Br~$\alpha$ to MIR sources} \label{sec:results:1}

\begin{figure}
\centering
\includegraphics[width=0.9\columnwidth]{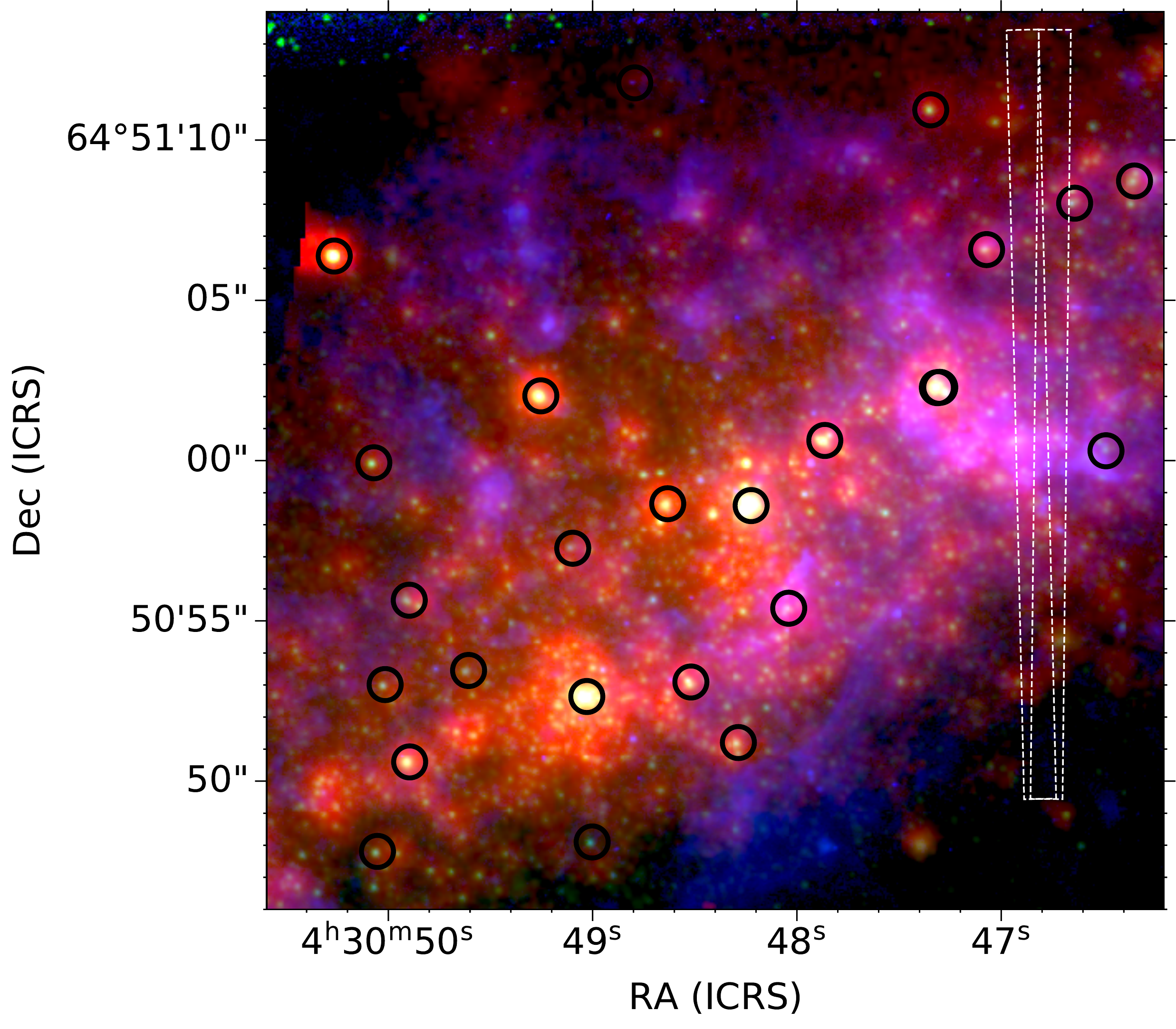}
\caption{Registration of the slits (white boxes) with GAIA sources (black circles). The background three-color image shows archival HST F656N (blue) and F814W (green) along with the KL-band (red). The astrometric solution is accurate to $\sim0\farcs1$\label{fig6}}
\end{figure}

Imaging of the KL-band continuum, which for simplicity we also refer to as ``2$\mu$m" continuum, was obtained with the NIRSPEC Slit-Viewing Camera (SCAM) during spectral exposures and used to determine accurate positioning of the slits on the sky (Fig.~\ref{fig6}). To perform astrometric calibration, we first combined sky-subtracted SCAM images for each of the two on-target positions, and measured the pixel positions of bright IR sources detected within each image. The IR sources were then matched to sources with well-established ICRS coordinates from the Gaia DR2 catalog \citep{gaia2016,gaia2018b}, as marked in Fig.~\ref{fig6}.  Residuals of the best-fit astrometric solutions are $\sim0\farcs1$ (RMS). 

The SCAM imaging indicates a seeing FWHM of $0\farcs8\simeq13$ pc, which should be capable of resolving separate continuum emission peaks from MIR1 and MIR2, which are separated by $1\farcs5$. MIR1 indeed coincides with a local peak in the KL continuum, but one that is fainter than continuum to its east and especially near cluster \#10 to the NE (the brightest KL-band peak in Fig.~\ref{fig2}, right). Moreover MIR2, despite its bright 10$\mu$m continuum, has no clear KL-band counterpart. Radio continuum is detected at MIR2 as a southern extension with no clearly detected source. Considering that MIR2 is the brightest 10$\mu$m continuum sources in NGC~1569, its lack of both radio and NIR continuum emission suggests suggests that it hosts extremely dense {\hii} regions and could be in a very early phase of formation, as proposed by \citet{tokura2006}. 

The morphology of {\bralpha} emission is illustrated by the echelle spectrum and intensity profile in Fig.~\ref{fig3}. The brightest {\bralpha} source is identified as ``Br-MIR1'' because of its clear association with the compact MIR sources. Br-MIR1 encompasses a sharp peak at the location of MIR1 and prominent shoulder of emission extending $\sim2{\arcsec}$ to its south, past the expected position of MIR2. The size of Br-MIR1, calculated from a single Gaussian fit to its intensity profile, is $\Delta\theta_{\rm FWHM}\simeq2\farcs2\simeq35$ projected-pc. Given the $\sim$1\farcs5 separation between MIR1 and MIR2, any detectable emission from MIR2 should be resolved by NIRSPEC. Rather, Br-MIR1 exhibits only the peak at MIR1, suggesting faint or negligible {\bralpha} emission from MIR2. This structure appears to closely follow the radio continuum (Fig.~\ref{fig1}). Despite no clear signal from MIR2, the shoulder extending from MIR1 to MIR2 presents a potential physical link between the two sources.  

The second {\bralpha} source identified in Fig.~\ref{fig3}, Br-N, is located 3$\arcsec$ to the north of Br-MIR1 and has best-fit size $\Delta\theta_{\rm FWHM}\simeq1\farcs2\simeq20$ projected-pc.
Br-N has no relation to MIR2 nor MIR1 but could be linked to the radio emission near cluster \#10 from \citet{wesmoquette2007b}. Further north, {\bralpha} emission is tentatively in narrow peaks up to $\sim$8{\arcsec} past Br-MIR1. These fluctuations could trace density enhancements like filaments intersecting the slit or possibly thin shells of ionized gas like those observed around many {\hii} regions \citep{sewlio2004}. 

\subsection{Ionized Gas Kinematics} \label{sec:results:2}

 Kinematic properties were measured from the spectra extracted for Br-MIR1 and Br-N with apertures of radius $r=7$pix$\simeq1\farcs1$
 (Fig.~\ref{fig4}). 
 The {\bralpha} lines from both sources appear remarkably symmetric; $\chi^{2}_{\nu}=1.6$ and 1.2 for a best-fit Gaussian profile to the spectrum of Br-MIR1 and Br-N, respectively. A two-component Gaussian profile provides a better model for Br-MIR1, however, to account for its broad wings ($\chi_{\nu}^2 = 1.3$; Fig.~\ref{fig4}). The measured properties of the {\bralpha} line, including both the single-component and two-component fit of Br-MIR1, are reported in Table~\ref{tab1}.

Br-MIR1 exhibits an extremely narrow line profile, only $\Delta V_{\rm FHWM}=42\pm2$ {\kms}. This result in confirmed by the {\siv} spectrum from TEXES, shown with its best-fit single Gaussian profile in Fig.~\ref{fig5} ($\chi^2_{\nu}=0.87$). Like {\bralpha}, the {\siv} line is symmetric and narrow, with a best-fit line width $\Delta V_{\rm [S\textsc{iv}]} = 41\pm2$. The centroid velocities of {\bralpha} and {\siv} are coincident to within $\lesssim5$ {\kms} ($-85\pm 1$ {\kms} and $-89\pm4$ {\kms}, respectively), suggesting both lines originate in the same gas. Taken together, the results leave little doubt of the quiescent nature of ionized gas across Br-MIR1, NGC 1569's giant embedded star forming region.
 
 \subsubsection{Two-component Structure} \label{sec:results:2:1}
 
 The {\bralpha} emission from Br-MIR1 exhibits two velocity components: a narrow core with line width (FWHM) $\Delta V_n=32\pm3$ {\kms} and a broader pedestal with $\Delta V_b= 74\pm12$. Each component contributes $\sim$50\% of the total line flux. A narrow+broad {\bralpha} line profile is common for embedded SSCs, however such massive clusters are often observed to have broad-component line widths of $\Delta V_b\sim250-350$ {\kms}\citep{henry2007,beck2008,cohen2018}, much larger than that of Br-MIR1. For comparison, the embedded supernebula in NGC~5253 exhibits a {\it narrow} component line width $\Delta V_n\sim75$ {\kms} \citep{cohen2018}. The line profile of Br-MIR1 is perhaps most comparable to Galactic hypercompact {\hii} regions, which have typical linewidths of 30-50 {\kms}\citep[e.g.,][]{hoare2007,sewilo2004}. We note that the broad component does not appear on the {\siv} line in Fig.~\ref{fig5}, possibly a reflection of nebular excitation structure.
 
\subsubsection{North-South Variation}

The centroid velocity of Br-MIR1 is consistent with the stellar and {\hi} gas systemic velocities for NGC 1569 \citep{johnson2012}, but shifts towards higher velocity to the north of Br-MIR1, with $V_{\rm hel}=-72$ {\kms} measured for Br-N. The implied gradient is roughly $+5$ {\varcs} or 0.3 {\vpc} from south to north. 
 
\begin{figure*}
\includegraphics[width=0.2\textwidth]{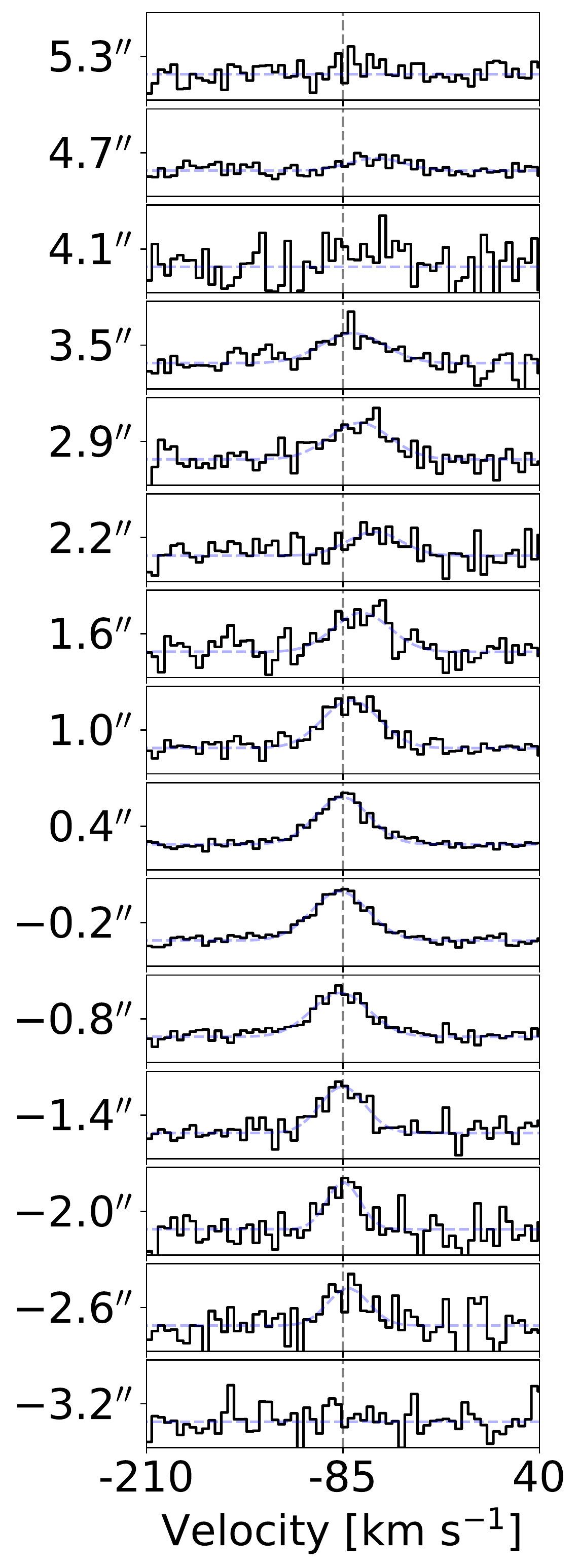}
\includegraphics[width=0.45\textwidth]{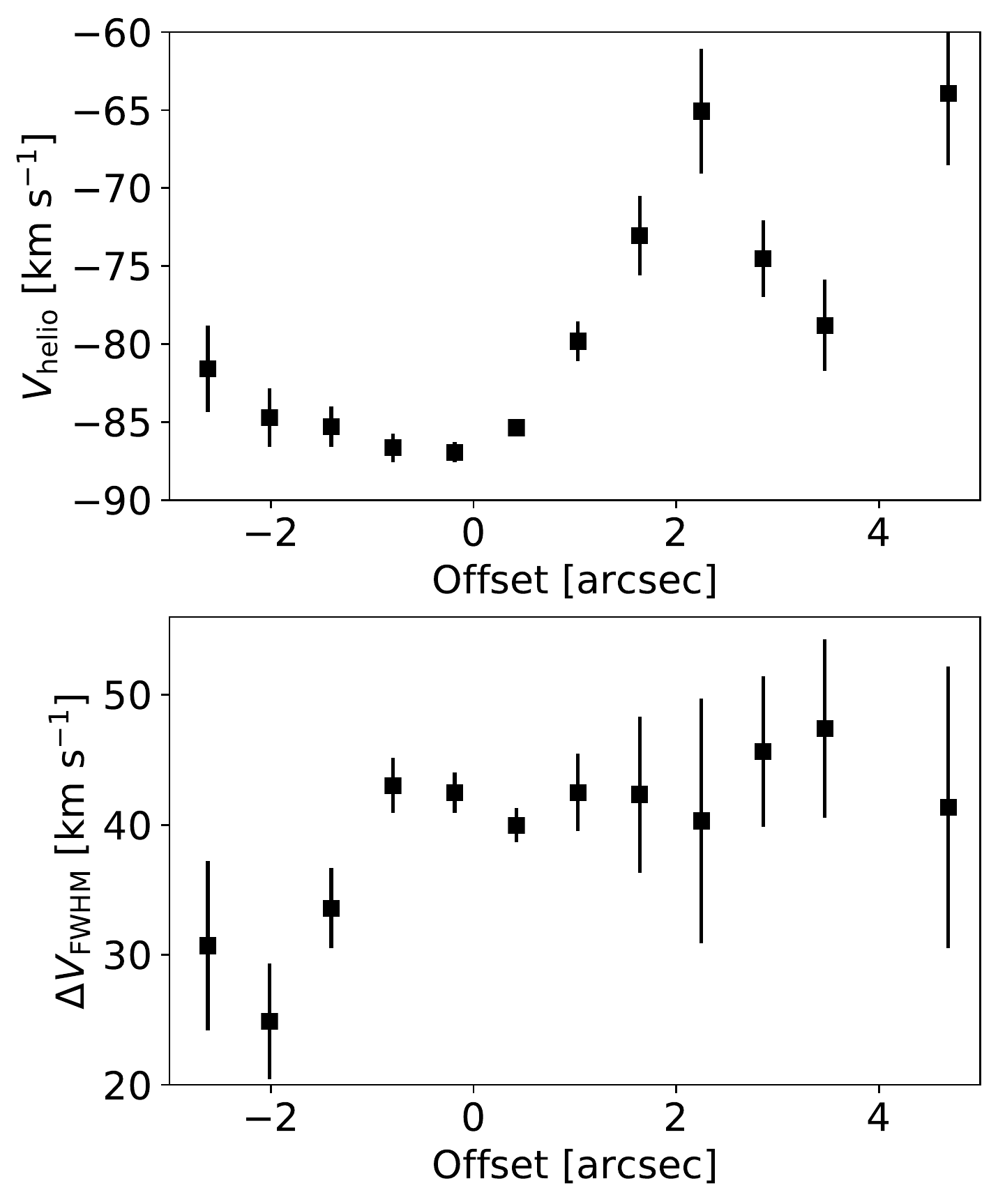}
\centering
\caption{ The Br $\alpha$ spectrum as a function of position along the slit (north-south). {\it Left}) Observed spectrum extracted every $0\farcs6=4$ pix along the slit axis (black line) and fit with a Gaussian profile (blue-dashed line) to measure velocity centroid and line width. The vertical black dashed line shows the systemic velocity, $V_{\rm sys}=-85$ {\kms}. {\it Right}) Centroid velocity (top) and line width (bottom) versus position along the slit (offset increases north).  \label{fig7} }
\end{figure*}

The spatial dependence of centroid velocity and line width along the north-south slit axis is investigated further by extracting 1D spectra from non-overlapping 4 pixel-wide ($0\farcs$6) apertures along the spatial axis of the 2D spectrum and fitting the line with single-component Gaussian profiles. A change in the line shape and peak position is visible by eye in the observed spectra in Fig.~\ref{fig7}, left. Best-fit $V_{\rm hel}$ and $\Delta V$ are shown as a function of offset in Fig.~\ref{fig7}, right, revealing this pattern in more detail. A positive velocity gradient of $\simeq2$ {\kms} arcsec$^{-1}$ extends to the south of the peak of Br-MIR1, along its shoulder. The slight redshift south along the shoulder corresponds to a sharp decrease in line width from $\sim45$ {\kms} at the peak to $\sim30$ {\kms} at an offset of $-$2{\arcsec}. 

A much steeper positive gradient of $\sim10$ {\kms} arcsec$^{-1}$ in velocity centroid occurs to the north of Br-MIR1, where it reaches a maximum $V_{\rm hel}\simeq -65$ {\kms} at an offset of just over $+$2{\arcsec}, and then reversing sign and decreasing until the location of Br-N where $V_{\rm helio}\simeq -75$ {\kms}. The minimum velocity centroid seemingly corresponds to a local maximum in the line width of 45 {\kms}. 

\begin{table*}
\caption{Properties of the two {\bralpha} sources identified in Fig.~\ref{fig2} and \ref{fig3}.  \label{tab1}}
\centering
\begin{threeparttable}
\begin{tabular}{cccccc} 
\hline
\hline
{Source} & {Offset\tnote{a}} & {$\Delta\theta_{\rm FWHM}$\tnote{b}}  & {$F_{\mathrm{tot}}${[$10^{-17}$ W m$^{-2}$]}\tnote{c}} & {$V_{\mathrm{hel}}${[\kms]}\tnote{d}} & {$\Delta V${[\kms]}\tnote{e}} \\
\hline
\hline
Br-MIR1 & 0.0 & $2.2\pm0.2$ & $20\pm4$  & $-85\pm1$ & $42\pm2$  \\ 
Br-N & 3.1 & $1.2\pm0.3$ & $4\pm1$  & $-73\pm2$ &  $45\pm5$\\ 
\hline
\hline
{$F_{n}/F_{\mathrm{tot}}$\tnote{f}} & {$V_{\rm n}${[\kms]}\tnote{g}} & {$\Delta V_{\rm n}${[\kms]}\tnote{h}} & {$F_{b}/F_{\mathrm{tot}}$\tnote{i}} & {$V_{\rm b}${[\kms]}\tnote{j}} & {$\Delta V_{\rm b}${[\kms]}\tnote{k}} \\
\hline
\hline
0.55 &  $-85\pm1$ & $32\pm3$ & 0.45 & $-88\pm3$ & $74\pm 12$ \\
\dotfill & \dotfill & \dotfill &  \dotfill & \dotfill & \dotfill \\ 
\hline
\end{tabular}
\begin{tablenotes}[para]
\centering
\footnotesize
\item [(a)] Spatial offset (along slit, north is positive) from the brightest source, Br-MIR1 (Fig.~\ref{fig3}).
\item [(b)] Angular size of source (FWHM) from the best-fit single Gaussian profile to the 1D spatial profile (Fig.~\ref{fig2}). 
\item [(c)] {\bralpha} line flux derived from the best-fit model of the 1D spectra (Fig.~\ref{fig3}) extracted within a $2\farcs2$ aperture; for Br-MIR1, the total flux of both fit components.
\item [(d)] Heliocentric centroid velocity from a single-component fit of the {\bralpha} emission (Fig.~\ref{fig3}).
\item [(e)] {\bralpha} velocity line width (FWHM) of the single-component fit to the spectrum.
\item [(f)] Fraction of total flux in the narrow component from the best-fit two-component spectrum (Fig.~\ref{fig3}).
\item [(g)] Heliocentric centroid velocity of the narrow fit component.
\item [(h)] {\bralpha} velocity line width (FWHM) of the narrow fit component (Fig.~\ref{fig3}).
\item [(i)] Fraction of total flux in the broad component from the best-fit two-component spectrum (Fig.~\ref{fig3}).
\item [(j)] Heliocentric centroid velocity of the broad fit component.
\item [(k)] {\bralpha} velocity line width (FWHM) of the broad fit component.
\end{tablenotes}
\end{threeparttable}
\end{table*}

\section{Discussion: Properties of the Embedded Clusters} \label{sec:dis}

The sources MIR1 and MIR2 discovered in the high resolution mid-infrared imaging of \citet{tokura2006} are the first indications of the presence of heavily embedded young SSCs in the region, among the very youngest known.
Informed by lower resolution investigations of the host region, we can use our high-resolution VLA radio continuum data, along with our \bralpha\ fluxes and linewidths, to improve on the analysis done by Tokura et al. and characterize the 
{\hii} region that these two sources represent. 

\subsection{The host star-forming region}\label{sec:dis:1}

Much of our current understanding of the MIR sources is based on lower resolution mapping across their host region. In the radio, the region corresponds to a bright concentration of thermal free-free continuum powered by thousands of newborn massive stars \citep[e.g.,][]{lisenfeld2004,westcott2018,hindson2018}. The 6~cm maps from \citet{lisenfeld2004} and \citet{westcott2018} indicate a peak brightness of $\sim10$ mJy per 6$\arcsec$ beam, with $\gtrsim60\%$ of the emission being thermal. This number is comparable to the total observed flux of 2-3 mJy for MIR1 and MIR2 from the high resolution VLA data (Fig.~\ref{fig1}), suggesting these sources contribute the bulk of ionization of the region.

In addition to its bright radio continuum, the {\hii}~2 region is associated with a concentration of dust detected in 5--500 $\mu$m imaging \citep[e.g.,][]{lisenfeld2002,galliano2003,suzuki2018,lianou2014,mccormick2018}. Maps from Spitzer and the Herschel PACS instrument, with resolutions of $\sim$2--10$\arcsec$, clearly locate warm dust to the position of MIR1/MIR2, also corresponding to the peak reddening $E(B-V)$ inferred from NIR observations \citep[e.g.,][]{pasquali2011}. The coincidence of MIR1 and MIR2 with this knot of dust indicates that the sources are embedded.

\subsection{MIR1 and MIR2 as ionizing sources} \label{sec:dis:2}

\citet{tokura2006} estimate an infrared luminosity of $L_{\rm IR}=(2\mbox{--}10)\times 10^7$~{\lsun} for MIR1 and $L_{\rm IR}=(4\mbox{--}31)\times 10^8$~{\lsun} for MIR2, which we have scaled to a distance of $D=3.4$ Mpc. MIR1 is well-constrained by its detectable radio continuum. While MIR2 lacks such radio emission, it should dominate in the FIR to sub-mm and can thus be constrained by existing low resolution data. Summing the 12-1200 $\mu$m global flux densities from SCUBA reported in \citet{lisenfeld2002} yields a global IR luminosity of $L_{\rm IR}\sim 10^9$~{\lsun}, putting an upper limit on the combined luminosity of MIR1 and MIR2. Alternatively we can adopt an upper limit of $L_{\rm IR}\sim 4\times 10^8$~{\lsun}, the (distance-scaled) dust luminosity over the full central starburst region from SED modelling in \citet{lianou2014} (see their Table 4). This value is comparable to the optically thick model luminosity of MIR2 from \citet{tokura2006} and consistent with their interpretation of this mysterious source.

For clusters of less than 5 Myr of age, the implied Lyman continuum rates are then 
$\log_{10}(N_{\rm LyC})=50.6\mbox{--}51.3$ for MIR1 and $\log_{10}(N_{\rm LyC})=52.0\mbox{--}52.8$ for MIR2. MIR2 is comparable to the supernebula in NGC 5253, and thus has a mass of roughly $3-5\times 10^5~\rm M_\odot$, and contains roughly 2000 O stars. To be consistent with these implied Lyman continuum rates, MIR1 should have a radio continuum flux of $0.4\mbox{--}2$ mJy and MIR2 $12\mbox{--}50$ mJy at wavelengths of 6~cm. The observed flux for MIR1 is 2 mJy, as expected. However, the flux of MIR2 is $<1$~mJy, at least a factor of 10 too low. This suggests that the turnover frequency for the free-free spectrum is higher than 6 cm. 

The Brackett emission, detected bright, extended ($\sim2\farcs2\simeq35$ pc) source Br-MIR1, should follow radio continuum fluxes and similarly bear witness to a lack of ionization in MIR2. Using the observed 6~cm fluxes, we follow \citet{ho1990} to predict the expected {\bralpha} fluxes, and compare to the observed value. Adopting $T_e=1.5\times10^4$ yields expected fluxes of $4\times10^{-16}$ W m$^{-2}$ for MIR1 and $\lesssim 6\times10^{-17}$ W m$^{-2}$ for MIR2. The actual observed flux is $F_{\mbralpha}=2\times10^{-16}$ W m$^{-2}$ (Table~\ref{tab1}). This value is consistent with the total radio-predicted flux of MIR1 and MIR2 after correcting for extinction and considering additional slit loss. MIR2 thus likely contributes $\lesssim10\%$ of the emission of Br-MIR1, and gas belonging to MIR1 should dominate the line profile (Fig.~\ref{fig4}).  
 
Following \citet{tokura2006} and using higher frequency radio continuum fluxes, we can use the missing radio continuum flux from MIR2 to estimate a turnover frequency, and thus an emission measure, EM = $\int n_e^2~d\ell$. If we assume that MIR2 should have 12 mJy of radio flux at 6~cm but a flux of only 0.2 mJy is observed, since the optical depth goes as $\nu^{-2.1}$ \citep{draine2011}, the implied turnover frequency is roughly 50 GHz.  The emission measure is thus $EM\sim 10^{10}~\rm cm^{-6}\, pc.$ This high emission measure puts MIR2 into the category of hypercompact HII regions. For a uniform density region, assuming a size of 3 pc typical of GCs and NGC 5253 \citep{turner2017,consiglio2017}, this would apply an rms density of $n_e\gtrsim 10^5~\rm cm^{-3}$. Since the ionized gas is probably clumped, and since the supernebula associated with MIR2 probably consists of multiple individual HII regions \citep{beck2008,cohen2018,silich2020}, the ionized gas densities are probably higher than $10^5~\rm cm^{-3}$. 

\subsection{Dynamical Constraints} \label{sec:dis:3}
 Theoretical work on forming massive clusters have found that a cluster's ability to survive feedback from its first generation of stars and to retain enough gas to form a second stellar generation is closely linked to its initial mass concentration $M/R$ \citep[e.g.,][]{krause2016,silich2018,elmegreen2018}. Using the {\bralpha} line width we can estimate the concentration of the clusters powering Br-MIR1, a lower limit which likely traces the lower density gas in MIR1. Observed IR hydrogen recombination line widths result from the convolution of thermal Doppler broadening $\Delta V_{\rm{thermal}}=\sqrt{k_B T_e / m_H}$ and non-thermal turbulent broadening due to gravity or feedback. Assuming $T_e=1.5\times10^4$ K, the single-component line width $\Delta V_{\rm{obs}}=42\pm2$ {\kms} corresponds to a turbulent width $\Delta V_{\rm{turb}} = 31\pm 3$ {\kms}, where we have also subtracted the instrumental line width in quadrature. If the turbulent width is solely due to gravity we can estimate the compactness using the virial theorem \citep{maclaren1988}: $M/R=190\Delta V_{\rm{turb}}^2\sim 2\times 10^5$ {\msun}~pc$^{-1}$, assuming a $1/r$ density profile. These values are consistent with kinematic measurements of of SSCs in NGC~5253, II Zw 40, He 2-10, the Antennae \citep[e.g.,][]{henry2007,beck2013,beck2015,turner2003,cohen2018}. We note that if we instead use the narrow, broad component line widths of the two-component fit (Fig.~\ref{fig3}), we estimate $M/R\sim 10^5$~{\msun} pc$^{-1}$ and $M/R\sim 10^6$~{\msun} pc$^{-1}$.
 
The mass concentration and observed size ($R_{cl}\sim1{\arcsec}=16$ pc) imply a cluster mass of $M_{cl}\sim 3\times10^6$ {\msun}.
 This is an upper limit to the mass of MIR1, since the radius is unlikely to be 15 pc. A more appropriate mass might be if $R=3~{\rm pc}$, a typical globular cluster radius, is assumed, which gives a mass of $6\times 10^5$~{\msun}. 
 Indeed, the extinction-corrected ionizing photon luminosity derived above results in a mass of $M_{cl}\sim3\times10^5$ {\msun}, which does not include any direct absorption of photons by dust. 

\section{Summary}

Using NIRSPEC we have observed the {\bralpha} 4.05~$\mu$m emission line across the embedded star-forming region in NGC~1569, at the location of the galaxy's brightest {\hii} region complex and two candidate SSCs MIR1 and MIR2. Presented alongside VLA radio continuum mapping and TEXES {\siv} 10.5 $\mu$m spectroscopy, the results give us novel insight into feedback in a potential globular cluster formation site:
\begin{enumerate}
\item The recombination line flux of the {\hii} region complex is dominated by a bright source with size $\Delta\theta\simeq 2\farcs2=35$ pc, which we associate with the mid-IR source of \citet{tokura2006} MIR1 and hence identify as Br-MIR1. Its spatial profile is consistent with the radio continuum mapping and indicates a structure comprising a compact emission peak with a southern shoulder across the position of mid-IR source MIR2. The shoulder suggests a physical link between MIR1 and MIR2.

\item The ionized gas line width of the embedded {\hii} region complex is extremely symmetric and narrow, only 40-45 {\kms} FWHM, measured for both the {\bralpha} and {\siv} lines. This is comparable to Galactic hypercompact {\hii} regions. The flux and line width implies a mass of $\sim10^5$ within a $R_{cl}\sim1$ pc cluster. It appears that feedback from the super star clusters is currently ineffective at driving rapid gas dispersal due to their extreme youth and the high density of {\hii} regions within. 

\item The {\bralpha} and radio continuum fluxes of source MIR2 predicted by its MIR continuum \citep{tokura2006} are much larger than those observed. The ionized gas appears to be very dense, with ${\rm EM}\gtrsim 10^{10}~{\rm cm^{-6}~pc}$, such that MIR2 is in the extremely young hypercompact {\hii} region phase. 

\end{enumerate}

Clearly NGC~1569's embedded {\hii} region complex, hosting candidate SSCs MIR1 and MIR2, warrants further investigation. New high resolution maps of the dust and gas near these sources will be necessary for constraining their fundamental properties, determining their formation and evolution, and uncovering any physical link between them. In particular MIR2, the more massive SSC, is likely in the earliest phases of formation in which ionization is quenched in extremely dense {\hii} regions around thousands of individual massive stars. Such observational targets are going to be vital in understanding the formation of massive star clusters and their effects on the galaxies in which they reside.

\section*{Acknowledgements}

We thank the anonymous reviewer for providing thoughtful comments and suggestions which strengthened the paper. The data presented herein were obtained at the W. M. Keck Observatory, which is operated as a scientific partnership among the California Institute of Technology, the University of California and the National Aeronautics and Space Administration. The Observatory was made possible by the generous financial support of the W. M. Keck Foundation. The authors wish to recognize and acknowledge the very significant cultural role and reverence that the summit of Maunakea has always had within the indigenous Hawaiian community.  We are most fortunate to have the opportunity to conduct observations from this mountain.


\section*{Data availability}

The data underlying this article are publicly available through the Keck Observatory Archive at \url{https://www2.keck.hawaii.edu/koa/public/koa.php} under Program ID: U048 (PI: J.~Turner), along with the NRAO Science Data Archive at \url{https://archive.nrao.edu/archive/advquery.jsp} under Program ID: AT0227 (PI: J.~Turner).



\bibliographystyle{mnras}
\bibliography{masterbib} 






\bsp	
\label{lastpage}
\end{document}